\batchmode
\makeatletter
\def\input@path{{/home/mattias/Papers/Potential_2003//}}
\makeatother
\documentclass[american,aps,preprint,showpacs,amsfonts,amssymb]{revtex4}
\usepackage[T1]{fontenc}
\usepackage[latin1]{inputenc}
\usepackage{amsmath}
\usepackage{babel}

\makeatletter

\providecommand{\LyX}{L\kern-.1667em\lower.25em\hbox{Y}\kern-.125emX\@}

\makeatother

\begin{document}


\preprint{hep-th/0307179}

\preprint{DAMTP-2003-68}

\title{Inflationary Cosmologies from Compactification?}

\author{Mattias N.R. Wohlfarth}

\email{M.N.R.Wohlfarth@damtp.cam.ac.uk}

\affiliation{Department of Applied Mathematics and Theoretical Physics, Centre for Mathematical
Sciences, University of Cambridge, Wilberforce Road, Cambridge CB3 0WA, U.K.}

\begin{abstract}
We consider the compactification of \( (d+n) \)-dimensional pure gravity and
of superstring/M-theory on an \( n \)-dimensional internal space to a \( d \)-dimensional
FLRW cosmology, with spatial curvature \( k=0,\pm 1 \), in Einstein conformal
frame. The internal space is taken to be a product of Einstein spaces, each
of which is allowed to have arbitrary curvature and a time-dependent volume.
By investigating the effective \( d \)-dimensional scalar potential, which
is a sum of exponentials, it is shown that such compactifications, in the \( k=0,+1 \)
cases, do not lead to large amounts of accelerating expansion of the scale factor
of the resulting FLRW universe, and, in particular, not to inflation. The case
\( k=-1 \) admits solutions with eternal accelerating expansion for which the
acceleration, however, tends to zero at late times. 
\end{abstract}
\maketitle

\section{Introduction}

The relationship between cosmology and superstring or M-theory is, at least,
twofold. On the one hand, cosmology greatly benefits from the existence of an
underlying, and consistent, unified theory of fundamental particle interactions
and of gravity. The hope may be expressed that the cosmology of our universe
will one day become derivable from higher dimensions. On the other hand, it
is well known that high energy phenomena near the Planck scale, which are described
by string/M-theory, are almost inaccessible to local observations. Therefore,
cosmological observations, influenced by physics on an enormous range of scales,
become probably the most important testing ground for these theories.

One particular aspect of cosmology, based on recent observations of supernovae
and the microwave background, is the emerging evidence for an accelerating expansion
of the universe. Not only a present epoch of accelerating expansion is indicated,
but also an inflationary epoch in the distant past. Much attention has been
paid recently to the construction of cosmological solutions, from compactifications
of string/M-theory, which exhibit such behaviour. The possibility of having,
at least one period of, accelerating expansion of the FLRW scale factor has
first been demonstrated in \cite{ToWo03}, thereby contradicting the general
belief that this was impossible, which was motivated by the difficulties to
find maximally symmetric de Sitter solutions from compactifications of supergravity
theories and by a no-go theorem for such compactifications \cite{Gib85}. 

The work of \cite{ToWo03} was done in the context of pure gravity. Subsequently,
it was generalised to include fluxes, as are necessary to consider in string/M-theory
compactifications, and, furthermore, a connection to S-branes was established
\cite{Follow} (see references to the S-brane literature therein). The original
solution (in eleven dimensions), in fact, turns out to be the zero flux limit
of an electrically charged SM2-brane solution. An investigation into the properties
of the phase of accelerating expansion in these solutions, performed in \cite{Woh03},
revealed the fact that this phase is very short in the sense that it only allows
a fixed amount of expansion of the universe, independent of the parameters.
The expansion factor turned out to be about two, i.e., of an order of magnitude
\( \mathcal{O}(1) \), and a similar analysis suggests that this is not improved
in other examples \cite{OhtEF}. So the known solutions do not apply to inflationary
scenarios. They might, however, according to \cite{GKL03}, turn out to be useful
for describing present day acceleration.

An intuitive explanation for the short accelerating phase and the resulting
smallness of the expansion factor was given in \cite{EmGa03} from the four-dimensional
point of view (we consider \( d+n \) dimensions here). The compactification
produces scalar fields as moduli of the metric of the internal space. The effective
lower-, \( d \)-dimensional theory, written down for the case of a single scalar
\( \phi  \), is of the generic form
\begin{equation}
\label{eq. acsingle}
S=\int \sqrt{-g}\left[ \frac{1}{2}R-\left( \nabla \phi \right) ^{2}-2V(\phi )\right] .
\end{equation}
Considering FLRW solutions for the metric, i.e.,
\begin{equation}
\label{eq. FLRWmetric}
ds^{2}=-dt^{2}+a(t)^{2}d\Sigma ^{2}_{k}\, ,
\end{equation}
where \( k=0,\pm 1 \) denotes the curvature of the \( (d-1) \)-dimensional
spatial sections, it now follows from the equations of motion of the above action
that the scale factor can exhibit accelerating expansion only if the potential
of the scalar field is larger than its kinetic energy and satisfies,
\begin{equation}
\label{eq. accel}
V>\frac{d-2}{2}\dot{\phi }^{2}\, .
\end{equation}
 In the cosmological compactifications discussed above, the scalar field starts
out at infinite value but with high kinetic energy. It runs into a positive
and steep exponential potential, turns around and falls back. At the turning
point, the solutions are dominated by potential energy and, consequently, the
scale factor accelerates. But this phase cannot last long because of the steepness
of the exponential potential.

The acceleration condition (\ref{eq. accel}) is easily extended to the case
of several scalar fields by summing their kinetic energies on the right hand
side. It presents a very simple criterion that has to be satisfied in order
to find solutions with accelerating expansion. Note that a necessary requirement
for this purpose is to have positive potentials. 

An almost too easy way to satisfy the acceleration condition, which we do not
further discuss here, is opened up by so called phantom scalar fields, see e.g.
\cite{Phantom}, motivated perhaps from the starred string theories that are
obtained from the conventional ones by timelike T-duality \cite{Star}. These
phantom fields have a different sign in their kinetic energy term, and, replacing
\( \dot{\phi }^{2}\mapsto -\dot{\phi }^{2} \), one finds that the acceleration
condition is always satisfied when the potential is positive and possibly also
when it is negative.

Another way of extending the known results on accelerating cosmologies has been
followed in \cite{CHNW03,CHNOW03}, where the authors find new solutions by considering
compactifications on product spaces, in which each factor is provided with its
own volume modulus. They do not find improved properties of the accelerating
phase, but they make an important observation. So far the discussion has assumed
\( k=0 \). If negative curvature is introduced on the spatial sections of the
cosmology, a solution with critical scale factor, i.e., with \( a\sim t \),
can be found and a perturbation around this solution may lead to eternal acceleration. 

The plan of this paper is to embed all the results and observations on accelerating
cosmologies obtained from compactifications of string/M-theory into a single
coherent point of view. To achieve such an understanding, we consider the effective
lower-dimensional theories and investigate the generic properties of the resulting
potential for the scalar fields. We do not have to solve equations of motion,
and do not present any new solutions, but our discussion will show clearly which
types of solutions may exist. Highlighting examples are taken from previous
papers. A question of particular interest will be whether there are solutions
with inflationary behaviour.

Section \ref{sec. attractive} considers attractor solutions for the scale factor
of an FLRW cosmology in the system of gravity coupled to a single scalar field
moving in an exponential potential. These are related to the characteristic
exponent of the potential. We argue that the case of many scalar fields can
effectively be reduced to the single scalar case and calculate the characteristic
exponents for more complicated potentials. The basic dimensional reduction of
\( (d+n) \)-dimensional gravity is performed in section \ref{sec. prodspace}.
The potential for the scalar fields is determined, and its consequences for
possible cosmologies are discussed. Section \ref{sec. MCompact} extends this
discussion to string/M-theory compactifications by including relevant fluxes
and the dilaton field. Section \ref{sec. truncate} presents the mechanism of
consistent truncation which reduces the number of scalar fields and might possibly
improve on our results. We conclude with a discussion in section \ref{sec. discuss}.

\section{Attractor Solutions\label{sec. attractive}}

Before compactifying \( (d+n) \)-dimensional pure gravity or string/M-theory
on product spaces down to \( d \) dimensions, as we will do in sections \ref{sec. prodspace}
and \ref{sec. MCompact}, let us first discuss a few features of the arising
lower-dimensional theory. For the scalar fields we will find potentials that
are sums of exponential terms. It will, however, turn out that, in order to
understand the case of many scalar fields, it is sufficient to understand the
case of a single scalar field. So let us for now restrict our attention to the
action (\ref{eq. acsingle}) and consider an exponential potential
\begin{equation}
V(\phi )=\Lambda e^{-2\alpha \phi }\, .
\end{equation}
When the potential is positive with \( \Lambda >0 \), the system of gravity
coupled to such a scalar field exhibits cosmological late-time attractor solutions,
see e.g. \cite{Attract}. For the metric assume the familiar \( d \)-dimensional
FLRW cosmologies as in (\ref{eq. FLRWmetric}). The evolution of the scale factor
\( a(t) \) then approaches a power law behaviour \( a(t)\sim t^{\gamma } \),
where the exponent \( \gamma  \) is related to the characteristic exponent
\( \alpha  \) appearing in the potential. Combining the four-dimensional classification
of \cite{Hal87} with the \( d \)-dimensional analysis of \cite{Tow01} (for
\( k=0 \)), one finds the attractor solutions given in table \ref{tab. relations}.
The different ranges of \( \alpha ^{2} \) are divided by the `critical' and
`hypercritical' characteristic exponents\begin{subequations}
\begin{eqnarray}
\alpha _{c}^{2} & = & \frac{2}{d-2}\, ,\\
\alpha _{h}^{2} & = & \frac{2(d-1)}{d-2}\, ,
\end{eqnarray}
\end{subequations}respectively.
\begin{table}
{\centering \begin{tabular}{c||c|c|c|c}
\( k\ddots \, \alpha ^{2} \)&
\( 0<\alpha ^{2}<\alpha _{c}^{2} \)&
\( \alpha ^{2}=\alpha _{c}^{2} \)&
\( \alpha _{c}^{2}<\alpha ^{2}<\alpha _{h}^{2} \)&
\( \alpha _{h}^{2}\leq \alpha ^{2} \)\\
\hline 
\hline 
\( -1 \)&
\( t^{(\alpha _{c}/\alpha )^{2}} \)&
\( t \)&
\( t \)&
\( t \)\\
\hline 
\( 0 \)&
\( t^{(\alpha _{c}/\alpha )^{2}} \)&
\( t \)&
\( t^{(\alpha _{c}/\alpha )^{2}} \)&
\( t^{(\alpha _{c}/\alpha _{h})^{2}} \)\\
\hline 
\( +1 \)&
\( t^{(\alpha _{c}/\alpha )^{2}} \)&
{*}&
{*}&
{*}\\
\end{tabular}\par}

\caption{\textit{\label{tab. relations}The scale factor for attractor solutions in
exponential scalar potentials for \protect\( k=0,\pm 1\protect \) and various
ranges of \protect\( \alpha ^{2}\protect \). The \protect\( k=+1\protect \)
cases marked with an asterisk undergo recollapse at late times and do not have
attractor solutions. Before that, they may, however, behave very much like the
\protect\( k=0\protect \) solutions. }}
\end{table}
 The phase space analysis of \cite{Hal87} also shows that the \( k=0 \) solutions
are unstable against perturbations in the spatial curvature of the universe.

Arbitrary amounts of accelerating expansion, or inflation, of the scale factor
\( a(t)\sim t^{\gamma } \), can only be obtained in two cases. In the standard
case, the exponent is \( \gamma >1 \) (or equivalently, \( \alpha ^{2}<\alpha _{c}^{2} \)):
the scale factor approaches an inflating attractor. The second case is the critical
one where \( \gamma =1 \), which is realised for \( \alpha ^{2}=\alpha _{c}^{2} \)
when \( k=0 \) and for \( \alpha ^{2}\geq \alpha _{c}^{2} \) when \( k=-1 \).
Here, accelerating expansion can be achieved by choosing initial conditions
such that the scalar fields start slowly rolling off the exponential potential.
Later, the solution approaches the critical attractor, and, although tending
to zero, the acceleration never stops. If, however, the attractor is a decelerating
one, as for \( \alpha ^{2}>\alpha _{c}^{2} \) and \( k=0 \), then the approach
of an actual solution, even setting initial conditions such that it starts off
to be accelerating, to this attractor can produce only a short phase of accelerating
expansion. It will quickly turn into decelerating expansion. 

Now return to the case of many scalar fields \( Q_{i} \) moving in a potential
that is a sum of positive exponential terms, i.e., given by
\begin{equation}
\label{eq. Vgeneric}
V(Q_{i})=\sum _{i}\Lambda _{i}e^{-2\sum _{j}\alpha _{ij}Q_{j}}
\end{equation}
where \( \Lambda _{i}>0 \). Consider the fields \( Q_{i} \) to depend on time
only, for an application to cosmological scenarios. Their time evolution is
governed by their equations of motion and the Friedmann constraint which follows
from the Einstein equations,\begin{subequations}
\begin{eqnarray}
\ddot{Q}_{i}+(d-1)H\dot{Q}_{i}+\frac{\partial V}{\partial Q_{i}} & = & 0\, ,\\
\alpha _{c}^{2}\sum _{i}\dot{Q}_{i}^{2}+2\alpha _{c}^{2}V(\mathbf{Q})-k\alpha _{h}^{2}a^{-2} & = & (d-1)H^{2}\, ,\label{eq. Friedmann} 
\end{eqnarray}
\end{subequations}where the quantity \( H=\dot{a}/a \) denotes the Hubble constant.
This time evolution generically moves the scalar fields far away from the origin
(\( Q_{i}=0 \)) in field space, at least when they are not trapped in a local
minimum of the potential. This fact allows us to reformulate the problem of
many scalar fields effectively in terms of a single one. At large norm squared
in field space, i.e., for \( \sum Q_{i}^{2}\gg 0 \), it is only a single exponential
term that dominates the potential \( V(Q_{i}) \). To this dominating term,
labelled by \( i \), say, corresponds a direction of steepest descent of the
potential, given by the vector \( \mathbf{q}_{(i)} \) with components \( \mathbf{q}_{(i)j}=\alpha _{ij} \),
and the fields will approximately evolve into this gradient direction at late
times. So going to infinity in field space along this direction, precisely as
\( \mathbf{Q}=|\mathbf{Q}|\frac{\mathbf{q}_{(i)}}{|\mathbf{q}_{(i)}|} \) with
\( |\mathbf{Q}|\rightarrow \infty  \), provides effective characteristic exponents
for an exponential potential of multiple scalar fields. Substituting the above
relations,
\begin{equation}
\Lambda _{i}e^{-2\sum _{j}\alpha _{ji}Q_{j}}\sim \Lambda _{i}e^{-2|\mathbf{q}_{(i)}||\mathbf{Q}|}\, ,
\end{equation}
one finds their values to be
\begin{equation}
\label{eq. chex}
\alpha _{i}^{2}=|\mathbf{q}_{(i)}|^{2}=\sum _{j}\alpha _{ij}^{2}\, .
\end{equation}
This argument will be very useful later in discussing the possible types of
attractor solutions occurring in product space compactifications.

This discussion of characteristic exponents for complicated scalar potentials
also nicely complements the phase space analyses that have been done previously,
albeit for simpler potentials. Special cases include, for instance, the `cross-coupling'
exponential potential, investigated in \cite{GPCZ03}, where further references
may be found.

\section{Compactifications on Product Spaces\label{sec. prodspace}}

We set out, in this section, to study compactifications of \( (d+n) \)-dimensional
gravity on product spaces. Extensions to include fluxes and dilaton fields appropriate
to compactifications of string/M-theory will be studied in the following section
\ref{sec. MCompact}, where a part of the material produced here will be needed.

\subsection{Dimensional Reduction}

The action of \( (d+n) \)-dimensional pure Einstein gravity on a spacetime
with metric \( g_{MN} \) is completely determined by the Ricci scalar \( R \)
of \( g_{MN} \),
\begin{equation}
\label{eq. action}
S_{(d+n)}=\int _{(d+n)}\sqrt{-g}\, R\, .
\end{equation}
The metric is split into its \( d \)-dimensional and its \( n \)-dimensional
internal part according to
\begin{equation}
\label{eq. ansatz}
ds^{2}=\tilde{g}_{\mu \nu }(x)dx^{\mu }dx^{\nu }+\hat{g}_{mn}(x,y)dy^{m}dy^{n}\, ,
\end{equation}
where the internal metric may also depend on the \( d \)-dimensional coordinates.
This coordinate-dependence, however, is realised in a particular way. The internal
space is taken to be a product of Einstein spaces \( E_{p}(\Lambda ) \), characterised
by their dimension \( p \) and a constant curvature \( \Lambda  \) that can
have either sign or be zero. Each factor space is multiplied by an \( x \)-dependent
function, so that its volume becomes \( x \)-dependent. These are the only
moduli fields kept of the internal space metric. Otherwise, the factor Einstein
spaces are assumed to be arbitrary but \( d \)-spacetime-independent. The sum
of all dimensions adds to \( \sum p_{i}=n \). Then the internal space metric
and the Ricci tensor are block-diagonal and, in suitable conventions, they are
given by\begin{subequations}\label{eq. prodspace}
\begin{eqnarray}
d\hat{s}^{2} & = & \sum _{i}e^{2F_{i}(x)}\hat{\omega }_{(i)\alpha \beta }(y)dy_{(i)}^{\alpha }dy_{(i)}^{\beta }\, ,\\
\hat{R}_{(i)\alpha \beta } & = & \Lambda _{i}(p_{i}-1)\hat{\omega }_{(i)\alpha \beta }\, .
\end{eqnarray}
 \end{subequations}One clear advantage of such a configuration is that the Einstein
equations with mixed indices in the \( d \)- and \( n \)-dimensional spacetime
parts, \( R_{\mu m}=0 \), are automatically satisfied. This is the case because
\( R_{\mu m} \) can be expressed in terms of \( x \)-derivatives of the internal
space Christoffel symbols \( \hat{\Gamma }^{p}_{mn} \), but these turn out
to be \( x \)-independent. 

Substituting the metric ansatz (\ref{eq. ansatz}) into the \( (d+n) \)-dimensional
action (\ref{eq. action}) causes a reduction down to \( d \) dimensions (which
is consistent as will be explained below),
\begin{equation}
\label{eq. action4}
S_{d}=\int _{d}\sqrt{-\tilde{g}}\, e^{\Sigma }\left[ \tilde{R}-2\tilde{\square }\Sigma -\left( \tilde{\nabla }\Sigma \right) ^{2}-\sum p_{i}\left( \tilde{\nabla }F_{i}\right) ^{2}+\hat{R}\right] ,
\end{equation}
where \( \Sigma (x) \) is defined as the sum of the volume moduli weighted
by the dimension of their respective factor spaces,
\begin{equation}
\label{eq. Sigma}
\Sigma (x)=\sum p_{i}F_{i}(x)\, .
\end{equation}
Due to the special form of the internal space, its Ricci scalar depends only
on \( x \), as
\begin{equation}
\label{eq. Ricciintern}
\hat{R}=\sum e^{-2F_{i}}\Lambda _{i}p_{i}(p_{i}-1)\, .
\end{equation}
 Note also that the \( x \)-independent integral over the volume of the internal
space,
\begin{equation}
\int _{n}\prod \sqrt{\hat{\omega }_{(i)}}\, ,
\end{equation}
has been removed by an appropriate redefinition of the \( d \)-dimensional
Newton constant. After dimensional reduction, however, the gravity part of \( S_{d} \)
is not in the standard Einstein-Hilbert form: \( \tilde{R} \) is multiplied
by an additional function \( e^{\Sigma } \). So in order to avoid possible
interpretational problems with a spacetime-varying Newton constant, one has
to remove \( e^{\Sigma } \). This transformation into the Einstein frame is
achieved by a conformal transformation of the metric \( \tilde{g}_{\mu \nu } \).
Note that such a conformal transformation leaves the causal structure of a spacetime
unchanged but changes physical properties otherwise, as for example the acceleration
of the scale factor, see also \cite{ToWo03}. The Einstein frame metric \( g_{E\mu \nu } \)
is defined by
\begin{equation}
g_{E\mu \nu }(x)=e^{\frac{2}{d-2}\Sigma (x)}\tilde{g}_{\mu \nu }(x)\, ,
\end{equation}
and transforming all terms of the action \( S_{d} \) leads to the Einstein
frame action
\begin{equation}
\label{eq. actionE}
S_{E}=\int _{d}\sqrt{-g_{E}}\left[ R_{E}+\frac{2}{d-2}\square _{E}\Sigma -\frac{1}{d-2}\left( \nabla _{E}\Sigma \right) ^{2}-\sum p_{i}\left( \nabla _{E}F_{i}\right) ^{2}+\hat{R}e^{-\frac{2}{d-2}\Sigma }\right] ,
\end{equation}
where \( \Sigma  \) and \( \hat{R} \) are given by (\ref{eq. Sigma}) and
(\ref{eq. Ricciintern}), respectively. The surface term \( \square _{E}\Sigma  \)
can now be dropped. This would not have been allowed before, in equation (\ref{eq. action4}),
since additional non-surface terms arise from it in the conformal transformation.

This compactification of the original \( (d+n) \)-dimensional action on the
\( n \)-dimensional product space is consistent. This means that any solution
of the equations of motion derived from \( S_{E} \) (i.e., any solution of
the Einstein equations with the appropriate scalar energy-momentum and of the
equations of motion for the volume scalars \( F_{i} \)) also solves the Einstein
equations derived from \( S_{(d+n)} \). Hence, any four-dimensional solution
obtained can be lifted directly to higher dimensions.

\subsection{The Scalar Potential}

As shown in the previous subsection, the product space compactification (\ref{eq. prodspace})
produces the four-dimensional action \( S_{E} \) (\ref{eq. actionE}) in Einstein
frame where gravity is coupled to the volume scalars \( F_{i} \). In this action,
however, the kinetic terms of the scalar fields are not in standard form, which,
in our conventions, means that the system of gravity coupled to a single scalar
\( \phi  \) should have the action (\ref{eq. acsingle}). With an obviously
unproblematic change of the overall normalisation, \( S_{E} \) can be rewritten
as
\begin{equation}
S_{E}=\int _{d}\sqrt{-g_{E}}\left[ \frac{1}{2}R_{E}-\nabla _{E}\mathbf{F}^{T}\, \mathcal{P}\, \nabla _{E}\mathbf{F}+\frac{1}{2}e^{-\frac{2}{d-2}\Sigma }\sum e^{-2F_{i}}\Lambda _{i}p_{i}(p_{i}-1)\right] ,
\end{equation}
where \( \mathbf{F} \) is the column vector formed from the fields \( F_{i} \)
and the matrix \( \mathcal{P} \) is defined by its components
\begin{equation}
\label{eq. pdef}
\mathcal{P}_{ij}=\frac{1}{2(d-2)}p_{i}p_{j}+\frac{1}{2}p_{i}\delta _{ij}\, .
\end{equation}
As \( \mathcal{P} \) is symmetric, it can be diagonalised by an orthogonal
matrix \( \mathcal{S} \), satisfying \( \mathcal{SS}^{T}=\mathcal{S}^{T}\mathcal{S}=\openone  \),
such that \( \mathcal{SPS}^{T}=\textrm{Diag}(P_{i}) \) where \( P_{i} \) are
the eigenvalues of \( \mathcal{P} \). Several properties of \( \mathcal{P} \)
are needed for later calculations. The most important ones are\begin{subequations}
\begin{eqnarray}
\det \mathcal{P} & = & 2^{-(M-1)}(1+\frac{1}{d-2}\sum p_{i})\prod p_{j}\, ,\\
(\mathcal{P}^{-1})_{ii} & = & \frac{2}{p_{j}}-\frac{2}{d+n-2}\, ,
\end{eqnarray}
\end{subequations}where \( M \) is the number of scalar fields \( F_{i} \),
i.e., the number of factor spaces of the internal product space. Defining new
scalars \( Q_{i} \) from the old ones \( F_{i} \) by the invertible relation
\begin{equation}
\label{eq. relQF}
Q_{i}\equiv \sum _{j}\sqrt{P_{i}}\mathcal{S}_{ij}F_{j}\, ,
\end{equation}
the action is cast into the canonical form (henceforth the index \( _{E} \)
denoting the Einstein frame will be dropped)
\begin{equation}
\label{eq. actionQ}
S=\int _{d}\sqrt{-g}\left[ \frac{1}{2}R-\nabla \mathbf{Q}^{T}\nabla \mathbf{Q}-2V(\mathbf{Q})\right] .
\end{equation}
Expressed in the fields \( Q_{i} \), the scalar potential is given by
\begin{equation}
\label{eq. potential}
V(\mathbf{Q})=-\frac{1}{4}\sum _{i}\Lambda _{i}p_{i}(p_{i}-1)e^{-2\frac{2}{p_{i}}\sum _{j}\sqrt{P_{j}}\mathcal{S}_{ji}Q_{j}}\, .
\end{equation}
Positive contributions to the potential arise from negatively curved Einstein
(factor) spaces where \( \Lambda _{i}<0 \) and vice versa.

Now assume the existence of an extremum of the potential at a point \( \mathbf {Q}_{0} \)
in field space where \( \partial V/\partial Q_{i}=0 \) for all values of \( i \).
A short calculation shows
\begin{equation}
\label{eq. Vext}
\sum _{i,j}\sqrt{P_{i}}\mathcal{S}_{ij}\frac{\partial V}{\partial Q_{i}}=-2\left( 1+\frac{n}{d-2}\right) V\, .
\end{equation}
Hence, if there is an extremum of the potential (\ref{eq. potential}), then
it can only be at points \( \mathbf{Q}_{0} \) where \( V(\mathbf{Q}_{0})=0 \).
This implies, in particular, that there is no positive extremum, and this fact
will be of importance below. The non-existence of a positive extremum of the
scalar potential in a supergravity compactification is also the essence of the
no-go theorem \cite{Gib85}.

\subsection{Effective Attractor Solutions}

Now that we have the precise form (\ref{eq. potential}) of the potential of
the canonically normalised scalar fields \( Q_{i} \), we can make use of the
general discussion of section \ref{sec. attractive} and calculate the corresponding
characteristic exponents and effective attractor solutions.

Let all \( N \) internal factors be Einstein spaces of zero or negative curvature.
Then \( V(\mathbf{Q})\ge 0 \) everywhere. This restriction serves a dual purpose.
On the one hand, it excludes most scenarios with a big crunch that occurs whenever
the left hand side in the Friedmann constraint (\ref{eq. Friedmann}) becomes
negative such that the evolution equations for the scalar fields break down
(there still may be big crunches for the \( k=+1 \) cosmologies). On the other
hand, this implies that the dominating terms in the potential are positive and
then the above arguments for attractor solutions can be applied: a potential
which is unbounded below does not allow for cosmological attractor solutions
(as in the single scalar case). 

If \( V(\mathbf{Q})\neq 0 \) initially, then time evolution will always move
the scalar fields to a large norm in field space, because the potential has
no positive extremum, as we have seen. There exist up to \( M \) directions
\( \{\mathbf{q}_{(i)}\, |\, k=1,2...\} \) of steepest descent of the potential,
corresponding to the domination of a single exponential term far from the field
space origin. These directions are read off from \( V(\mathbf{Q}) \) as
\begin{equation}
q_{(i)j}=\frac{2}{p_{i}}\sqrt{P_{j}}\mathcal{S}_{ji}\, ,
\end{equation}
compare (\ref{eq. Vgeneric}), and they determine the cosmological attractor
solutions. According to (\ref{eq. chex}), the characteristic exponents are
given by
\begin{equation}
\label{eq. charexp}
\alpha _{i}^{2}=\sum _{j}\left( \frac{2}{p_{i}}\sqrt{P_{j}}\mathcal{S}_{ji}\right) ^{2}.
\end{equation}
The question, of whether one gets large amounts of accelerating expansion, or
inflation, from the product space compactification presented above, now reduces
to a calculation of these characteristic exponents \( \alpha _{i} \). From
the definition of the matrix \( \mathcal{P} \) in (\ref{eq. pdef}) and using
the fact that it is positive definite, one finds
\begin{equation}
\mathcal{P}_{ii}=\sum _{j}\left( \sqrt{P_{j}}\mathcal{S}_{ji}\right) ^{2}=\frac{p_{i}}{2}(\frac{p_{i}}{d-2}+1)\, ,
\end{equation}
which can be rewritten to give the following simple formula:
\begin{equation}
\alpha _{i}^{2}=\alpha _{c}^{2}\frac{p_{i}+d-2}{p_{i}}\, .
\end{equation}
Note that the characteristic exponents always lie in the range between the critical
one and the hypercritical one, \( \alpha _{c}<\alpha _{i}<\alpha _{h} \) for
\( d>2 \). The relevant attractor solutions in the potential \( V(\mathcal{Q}) \)
(there are as many as there are internal Einstein spaces of strictly negative
curvature) can be read off from table \ref{tab. relations} to be \( a(t)\sim t \)
for \( k=-1 \) and \( a(t)\sim t^{(\alpha _{c}/\alpha _{i})^{2}} \) for \( k=0 \).
None of them is accelerating. So for \( k=0 \) one can at most obtain a short
phase of accelerating expansion by choosing suitable initial conditions as explained
in section \ref{sec. attractive}. This is the same mechanism as in most of
the previously discussed exact solutions. In the \( k=-1 \) case, choosing
the right initial conditions can lead to eternal acceleration.

We emphasise that the important case of power law inflation, which would be
implied by the existence of accelerating attractor solutions with \( \alpha _{i}^{2}\le \alpha _{c}^{2} \),
is not realised in simple product space compactifications of pure gravity.

\section{String/M-Theory Compactifications\label{sec. MCompact}}

To extend the application of the above results to product space compactifications
of the various superstring theories or M-theory, one has to include the fields
appearing in the corresponding low energy effective actions, as given by the
ten and eleven-dimensional supergravities: antisymmetric field strengths \( F_{p} \)
with, possibly, dilaton couplings and the dilaton scalar field \( \phi  \)
itself.

\subsection{Four Cases}

The new starting point is the \( (d+n) \)-dimensional bosonic action
\begin{equation}
S_{(d+n)}=\int _{(d+n)}\sqrt{-g}\left[ R-\frac{1}{2}(\nabla \phi )^{2}-\frac{1}{2p!}e^{a\phi }F_{p}^{2}-\frac{1}{2}m^{2}e^{-\frac{5}{2}\phi }\right] ,
\end{equation}
where one might also want to include the mass term of massive IIA supergravity
\cite{Rom86} by having \( m\neq 0 \). We do not consider possible Chern-Simons
terms for the fluxes and their potentials here, as we will realise the fluxes
by volume forms on certain subspaces below, which makes these terms irrelevant.
After dimensional reduction down to \( d \) dimensions, only a few different
cases have to be considered. Compatible with the symmetries of the FLRW cosmologies,
i.e., with the symmetries of a space with the topology \( \mathbb {R}\times \Sigma _{k} \),
there can exist the following field strength forms after compactification: 

\textit{(i)} \( F_{0} \) and \textit{(ii)} \( F_{d} \), 

\textit{(iii)} \( F_{1} \) with non-zero components only along the real time
direction, and 

\textit{(iv)} \( F_{d-1} \) with non-zero components only on the spatial sections
\( \Sigma _{k} \). 

All other fluxes with non-negligible effects on cosmological scales are excluded
(of course, this reasoning does not apply to local fields, for instance, electromagnetic
fields in \( d=4 \)). 

In the dual cases \textit{(i)} and \textit{(ii)}, an analysis of the \( d \)-dimensional
equations of motion, for gravity, the field strengths and the scalars arising
from the compactification process, shows that the field strengths \( F_{0} \)
and \( F_{d} \) act as, and thus can be replaced by, genuine potentials for
the scalar fields. They effectively become time-dependent cosmological constants.
This becomes apparent, for example, in \cite{Woh03}, where an \( F_{4} \)
arising from an M-theory compactification to four dimensions violates the four-dimensional
strong energy condition by providing a positive scalar potential. \textit{(i)}
More explicitly, forms \( F_{0} \) in \( d \) dimensions arise from \( F_{p} \)
having components only in the internal space. We realise such \( F_{p} \) as
volume forms with field strength parameter \( b_{0} \), i.e., we set \( F_{p}=b_{0}vol_{d,d+1...} \),
and denote this type of reduction by
\begin{equation}
F_{p}(d,d+1...)\leadsto F_{0}()\, .
\end{equation}
The reduced action in the \( d \)-dimensional Einstein frame then becomes
\begin{equation}
S=\int _{d}\sqrt{-g}\left[ \frac{1}{2}R-(\nabla \phi )^{2}-\nabla \mathbf{Q}^{T}\nabla \mathbf{Q}-2V(\phi ,\mathbf{Q})\right] ,
\end{equation}
and the scalar potential is given by
\begin{equation}
\label{eq. pot00}
V(\phi ,\mathbf{Q})=V(\mathbf{Q})+\frac{1}{4}b_{0}^{2}e^{2a\phi -\frac{2}{d-2}\Sigma (\mathbf{Q})-2p_{1}F_{1}(\mathbf{Q})}+\frac{1}{8}m^{2}e^{-5\phi -\frac{2}{d-2}\Sigma (\mathbf{Q})},\quad p_{1}=p\, ,
\end{equation}
where previous definitions (\ref{eq. Sigma}), (\ref{eq. relQF}) and (\ref{eq. potential})
have been used for \( \Sigma  \), the scalars \( F_{i}(\mathbf{Q}) \) and
the potential term \( V(\mathbf{Q}) \), respectively. The original dilaton
field \( \phi  \) has been canonically normalised by rescaling \( \phi \mapsto 2\phi  \).
The volume scalars \( F_{i} \) should not be confused with field strengths.
\textit{(ii)} The reduction
\begin{equation}
F_{p}(01...d-1|d,...)\leadsto F_{d}(01...d-1)
\end{equation}
gives the potential
\begin{equation}
\label{eq. pot44}
V(\phi ,\mathbf{Q})=V(\mathbf{Q})+\frac{1}{4}b_{0}^{2}e^{-2a\phi -\frac{2(d-1)}{d-2}\Sigma (\mathbf{Q})+2p_{1}F_{1}(\mathbf{Q})}+\frac{1}{8}m^{2}e^{-5\phi -\frac{2}{d-2}\Sigma (\mathbf{Q})},\quad p_{1}=p-d\, .
\end{equation}

The cases \textit{(iii)} and \textit{(iv)} are also dual. Each field strength
\( F_{1} \) aligned along the time direction is equivalent to an additional
scalar field \( \psi  \) via the relation \( F_{1}=d\psi  \), and spatial
\( (d-1) \)-forms give rise to scalars via \( \star F_{d-1}=F_{1}=d\psi  \).
The reduced action takes the form
\begin{equation}
S=\int _{4}\sqrt{-g}\left[ \frac{1}{2}R-(\nabla \phi )^{2}-\nabla \mathbf{Q}^{T}\nabla \mathbf{Q}-e^{C(\phi ,\mathbf{Q})}(\nabla \psi )^{2}-2V(\phi ,\mathbf{Q})\right] 
\end{equation}
(after normalising \( \phi  \) and \( \psi  \)), where the function \( C \)
appears as a \( d \)-dimensional `dilaton coupling'. \textit{(iii)} For the
reduction
\begin{equation}
F_{p}(0|d,d+1...)\leadsto F_{1}(0)
\end{equation}
one finds\begin{subequations}
\begin{eqnarray}
C(\phi ,\mathbf{Q}) & = & 2a\phi -2p_{1}F_{1}(\mathbf{Q}),\quad p_{1}=p-1\, ,\\
V(\phi ,\mathbf{Q}) & = & V(\mathbf{Q})+\frac{1}{8}m^{2}e^{-5\phi -\frac{2}{d-2}\Sigma (\mathbf{Q})}\, ,
\end{eqnarray}
\end{subequations}In case \textit{(iv)} where
\begin{equation}
F_{p}(1...d-1|d...)\leadsto F_{d-1}(1...d-1)\, ,
\end{equation}
 one obtains\begin{subequations}
\begin{eqnarray}
C(\phi ,\mathbf{Q}) & = & -2a\phi -\frac{4}{d-2}\Sigma (\mathbf{Q})+2p_{1}F_{1}(\mathbf{Q}),\quad p_{1}=p-d+1\, ,\\
V(\phi ,\mathbf{Q}) & = & V(\mathbf{Q})+\frac{1}{8}m^{2}e^{-5\phi -\frac{2}{d-2}\Sigma (\mathbf{Q})}\, .
\end{eqnarray}
\end{subequations}

To answer the question whether inflation can be obtained, the same arguments
as in the preceding sections can be applied. Note that (\ref{eq. Vext}) can
easily be generalised. It is also true here that there do not exist any positive
extrema of the potential. So we may look at dominant terms in the potential
far from the origin in field space. Consider first the term arising from massive
supergravity which is proportional to \( e^{-2(5\phi /2+\Sigma /(d-2))} \).
Without even taking the \( \mathbf{Q} \)-dependence into account, one sees
that the characteristic exponent \( \alpha _{m} \) is larger than the critical
one, \( \alpha _{m}>\frac{5}{2}>\alpha _{c} \) for \( d>2 \). 

In cases \textit{(i)} and \textit{(ii)}, there is another term in the potential,
arising from the field strength. The respective field space norms of the exponents can
be calculated from
\begin{equation}
\alpha _{F}^{2}=a^{2}+\frac{\zeta ^{2}}{4}\sum _{i}\left( \sum _{j}\frac{p_{j}}{\sqrt{P_{i}}}\mathcal{S}_{ij}(1+\xi \delta _{j1})\right) ^{2}
\end{equation}
where the parameters are given by \( (\zeta ,\xi )=(\alpha _{c}^{2},2/\alpha _{c}^{2}) \)
in case \textit{(i)} and \( (\zeta ,\xi )=(\alpha _{h}^{2},-2/\alpha _{h}^{2}) \)
in case \textit{(ii)}, respectively. The derivation uses the expression for
the inverse of \( \mathcal{P} \) in terms of its determinant and its adjoint.
One finds
\begin{equation}
\label{eq. alpF}
\alpha _{F}^{2}=a^{2}+\frac{\zeta ^{2}[-\xi ^{2}p_{1}^{2}+((d+n-2)\xi +2(d-2))\xi p_{1}+n(d-2)]}{2(d+n-2)}\, .
\end{equation}
As an important example, look at string theory compactifications with \( d=4 \)
and \( n=6 \). The value of the dilaton coupling is given by \( a=\frac{1}{2}(5-p) \)
where the degree \( p \) of the field strength form is \( p=p_{1} \) in case
\textit{(i)} and \( p=p_{1}-4 \) in case \textit{(ii)}. An evaluation of the
above expression then gives \( \alpha _{F}^{2}=7 \) independent of \( p_{1} \),
and for both cases. This is greater than the critical exponent \( \alpha _{c}=1 \).
The conclusion, therefore, is that there are no accelerating attractor solutions
for \( k=0,+1 \). For \( k=-1 \), one finds again the critical attractor with
zero acceleration. 

In cases \textit{(iii)} and \textit{(iv)} there are no further contributions
to the potential. The difference here is a non-standard structure of the kinetic
term of the field \( \psi  \), including a dilaton-coupling. (Such a coupling
cannot be removed by simple field redefinitions.) Here, the above arguments
might not be applicable in a straightforward way. It can be shown that actions
with such non-standard scalar fields may admit power-law solutions quite different
from those obtained when there are only scalars with standard kinetic terms.
This happens, for example, in the model
\begin{eqnarray}
S & = & \int \sqrt{-g}\left[ \frac{1}{2}R-(\nabla \phi )^{2}-(\nabla Q)^{2}-e^{-2(a_{0}\phi +a_{1}Q)}(\nabla \psi )^{2}\right. \nonumber \\
 &  & \quad \quad \quad \quad \quad \quad \quad \quad \quad \quad \quad \left. -2\Lambda e^{-2\alpha \phi }-\frac{1}{4}M^{2}e^{-2(a_{2}\phi +a_{3}Q)}\right] 
\end{eqnarray}
that arises as a special case of both, \textit{(iii)} and \textit{(iv)}. But,
in these cases, the scale factor \( a(t)\sim t^{\gamma } \) never has a solution
with \( \gamma >1 \). So the non-standard coupling of the scalar fields does
not make it easier to find accelerated expansion. This is, in fact, to be expected
because these terms provide additional kinetic energy and make it harder to
satisfy the acceleration condition (\ref{eq. accel}).

\subsection{Scale Invariance}

The supergravity actions including a dilaton field have a scale invariance which
survives the compactification process. As was noted in \cite{GiTo87}, the scalar
potential \( V(\phi ,Q_{i}) \) in the compactified theory can then be written
as a product. Identify \( Q_{0}\equiv \phi  \). We expect that
\begin{equation}
V(Q_{\mu })=V(Q_{0})V(Q_{i})
\end{equation}
and that the dependence of \( V(Q_{0}) \) on \( Q_{0} \) is purely exponential.
To see that this is the case, note that the \( (M+1) \) scalar fields \( Q_{\mu } \),
arising from the original higher-dimensional dilaton and from the volume moduli
of the internal space, are only defined up to an \( SO(M+1) \) rotation, since
this leaves their kinetic terms invariant. To achieve a product decomposition
of the potential, one generically has to perform such a rotation. 

We will now construct this rotation explicitly for the \( F_{p}(d,d+1...)\leadsto F_{0}() \)
and \( F_{p}(01...d-1|d...)\leadsto F_{d}(01...d-1) \) reductions, respectively
(setting the mass parameter \( m \) to zero). We define new fields \( \widetilde{Q}_{\nu } \),
rotating by a matrix \( \mathcal{T}\in SO(M+1) \), as
\begin{equation}
Q_{\mu }\equiv \sum _{v}\mathcal{T}_{\mu \nu }\widetilde{Q}_{\nu }\, .
\end{equation}
The scalar potential is given from (\ref{eq. pot00}) or, respectively, (\ref{eq. pot44})
as

\begin{eqnarray}
V(\widetilde{Q}_{\nu }) & = & -\frac{1}{4}\sum _{k}\Lambda _{k}p_{k}(p_{k}-1)e^{-2\sum _{\nu }\left( \sum _{j}\frac{2}{p_{k}}\sqrt{P_{j}}\mathcal{S}_{jk}\mathcal{T}_{j\nu }\right) \widetilde{Q}_{\nu }}\nonumber \\
 &  & +\frac{1}{4}b_{0}^{2}e^{-2\sum _{\nu }\left( \mp a\mathcal{T}_{0\nu }+\zeta \sum _{ij}\frac{p_{i}}{2\sqrt{P_{j}}}\mathcal{S}_{ji}(1+\xi \delta _{i1})\mathcal{T}_{j\nu }\right) \widetilde{Q}_{\nu }}\, ,\label{eq. Vtilde} 
\end{eqnarray}
where we use the parameters \( (\zeta ,\xi )=(\alpha _{c}^{2},2/\alpha _{c}^{2}) \)
and the upper sign for \( F_{p}(d,d+1...)\leadsto F_{0}() \), and \( (\zeta ,\xi )=(\alpha _{h}^{2},-2/\alpha _{h}^{2}) \)
and the lower sign, for the potential from the \( F_{p}(01...d-1|d...)\leadsto F_{d}(01...d-1) \)
reduction. To split off, from the potential, a common factor of the form \( e^{-2(...)\widetilde{Q}_{0}} \),
the condition
\begin{equation}
\pm a\mathcal{T}_{00}=\sum _{j}\left( \zeta \sum _{i}\frac{p_{i}}{2\sqrt{P_{j}}}\mathcal{S}_{ji}(1+\xi \delta _{i1})-\frac{2}{p_{k}}\sqrt{P_{j}}\mathcal{S}_{jk}\right) \mathcal{T}_{j0}
\end{equation}
has to be satisfied for all values of \( k=1,\, ...,\, M-1 \). This can be
done by setting\begin{subequations}
\begin{eqnarray}
\mathcal{T}_{00} & = & \pm \frac{(\zeta -\alpha _{c}^{2})n+\zeta \xi p_{1}-2}{2a}\gamma \, ,\\
\mathcal{T}_{j0} & = & \sum _{l}\sqrt{P_{j}}\mathcal{S}_{jl}\gamma \, .
\end{eqnarray}
\end{subequations}The normalisation constant \( \gamma  \) ensures that \( \mathcal{T}_{\mu 0} \)
are components of an \( SO(M+1) \)-matrix. The condition is \( \sum _{\mu }\mathcal{T}_{\mu 0}^{2}=1 \)
and is satisfied for
\begin{equation}
\gamma ^{2}=4a^{2}\left[ \left( (\zeta -\alpha _{c}^{2})n+\zeta \xi p_{1}-2\right) ^{2}+a^{2}\alpha _{c}^{2}n(d+n-2)\right] ^{-1}.
\end{equation}

Now that we have shown that the potential can be written in product form, it
is interesting to check the characteristic exponent of the factor \( V(\widetilde{Q}_{0}) \).
It is \( \alpha _{0}=\frac{d+n-2}{d-2}\gamma  \) and will be subcritical, \( \alpha _{0}^{2}\leq \alpha _{c}^{2} \),
if, and only if,
\begin{equation}
\left( (\zeta -\alpha _{c}^{2})n+\zeta \xi p_{1}-2\right) ^{2}\geq a^{2}\alpha _{c}^{2}(d+n-2)(2-n\frac{d-4}{d-2})\, .
\end{equation}
For string theory compactifications with \( d=4 \) and \( n=6 \), the value
of the dilaton coupling is \( a=\frac{1}{2}(5-p) \) where the degree \( p \)
of the field strength form is \( p=p_{1} \) in the first case and \( p=p_{1}-4 \)
in the second. One checks that the exponent is subcritical for \( p=p_{1}>2 \)
or, in the second case, for \( p-4=p_{1}\leq 3 \) (exactly critical for \( p=7 \)). 

This is potentially interesting because one might imagine the potential having
a flat direction, becoming constant near infinity, in the \( M \)-dimensional
field space of the \( \widetilde{Q}_{i} \). Then choosing initial conditions
such that all fields start at rest in this region, they would evolve towards
an accelerating attractor in direction of \( \widetilde{Q}_{0} \), while at
the same time rolling very slowly off the almost flat potential in the other
directions. This might yield a large amount of inflation and would at the same
time provide a mechanism to stop the inflationary epoch. To look for such flat
directions note that, by the same argument as employed above, a single exponential
term of the potential factor \( V(\widetilde{Q}_{i}) \) dominates at large
\( |\widetilde{\mathbf{Q}}| \). It is of the form \( e^{-2\widetilde{\alpha }\cdot \widetilde{\mathbf{Q}}} \)
and constant if, and only if, \( |\widetilde{\alpha }|^{2}=0 \). The expression
\( V(\widetilde{Q}_{i}) \) is given by (\ref{eq. Vtilde}) with the only difference
that, in the exponents, one has to sum only over \( i \), i.e., over \( \nu \neq 0 \).
We calculate the characteristic exponents \( |\widetilde{\alpha }|^{2} \) for
all exponential terms in turn. For those in the sum over the curvatures \( \Lambda _{k} \)
of the internal space factors, we find
\begin{equation}
\widetilde{\alpha }_{k}^{2}=\frac{2}{d-2}+\frac{2}{p_{k}}-\gamma ^{2}\frac{(d+n-2)^{2}}{(d-2)^{2}}\, .
\end{equation}
 For the term arising from the field strength, using \( \alpha _{F}^{2} \)
from (\ref{eq. alpF}), one obtains
\begin{equation}
\widetilde{\alpha }_{F}^{2}=\alpha _{F}^{2}-\frac{1}{4}\gamma ^{2}(\alpha _{c}^{2}n+2)^{2}\, .
\end{equation}
For string theory compactifications down to four dimensions, these expressions
never become zero. Thus the potential does not have flat directions in \( \widetilde{Q}_{i} \)-space.

The arguments of this section also hold if one includes the mass term of IIA
supergravity but sets the field strength to zero. The calculation proceeds as
before. For the parameters one has to substitute \( a=-\frac{5}{2} \) and \( (\zeta ,\xi )=(1,0) \).
For \( d=4 \) and \( n=6 \), this case also allows a factorisation of the
potential into a pure exponential with characteristic exponent \( \alpha _{0}>1 \),
and into another factor that does not have any flat directions either.

\section{Consistent Truncations\label{sec. truncate}}

We have seen that the product space compactification produces characteristic
exponents that are always above the critical value \( \alpha _{c} \). As discussed
above, this results in the non-existence of solutions with genuine power-law
inflation. This section presents a mechanism of reducing the characteristic
exponents by truncating the number of scalar fields. 

Suppose the compactification produces two scalar fields \( \phi  \) and \( Q \).
In certain circumstances, it is then possible to truncate the compactified theory
consistently, such that only one scalar field \( \psi  \) is left. This is
performed by choosing a certain direction in the originally two-dimensional
field space. Set
\begin{equation}
\label{eq. trunc}
\phi =\frac{s}{\sqrt{1+\lambda ^{2}}}\psi \quad \textrm{and}\quad Q=\frac{\lambda }{\sqrt{1+\lambda ^{2}}}\psi 
\end{equation}
with \( s^{2}=1 \) and an at first arbitrary parameter \( \lambda  \) to be
determined later. (This procedure can be generalised to a higher number of fields
but becomes increasingly messy.) To see that such a truncation is interesting
and reduces the characteristic exponent in a contribution to the potential,
consider a typical term proportional to \( e^{-2(\alpha \phi +\beta Q)} \).
The characteristic exponent, before the truncation, is \( \alpha _{0}^{2}=\alpha ^{2}+\beta ^{2} \).
After the truncation, one finds
\begin{equation}
\label{eq. norm2}
\alpha _{\textrm{trunc}.}^{2}=\frac{\alpha ^{2}+\beta ^{2}\lambda ^{2}+2\alpha \beta s\lambda }{1+\lambda ^{2}}\, .
\end{equation}
This expression is smaller than \( \alpha _{0}^{2} \), unless \( \alpha s\lambda -\beta =0 \).
So, generically, a consistent truncation reduces the characteristic exponents,
and thus it presents another, possibly helpful mechanism of generating accelerating
expansion. 

To demonstrate this concept, we consider a reduction of the type
\begin{equation}
F_{4}(0123)\leadsto F_{4}(0123)
\end{equation}
in a \( (4+n) \)-dimensional theory. The internal space shall only have a single
factor and we set \( m=0 \). In terms of the fields \( \phi  \) and \( Q \),
the potential (\ref{eq. pot44}) becomes
\begin{equation}
V(\phi ,Q)=-\frac{1}{4}\Lambda n(n-1)e^{-2\sqrt{\frac{n+2}{n}}Q}+\frac{1}{4}b_{0}^{2}e^{-2a\phi -6\sqrt{\frac{n}{n+2}}Q}.
\end{equation}
The consistency condition for the truncation is that both equations of motion
for the scalar fields are solved simultaneously when substituting (\ref{eq. trunc}).
This determines
\begin{equation}
s\lambda =-\frac{a\sqrt{n(n+2)}}{2(n-1)}
\end{equation}
and
\begin{equation}
\frac{1}{2}\Lambda (n-1)=\frac{1}{2}b_{0}^{2}\left( \frac{a^{2}}{2(n-1)}+\frac{3}{n+2}\right) .
\end{equation}
Calculating the characteristic exponent \( \alpha ^{2}_{\textrm{trunc}.} \),
using (\ref{eq. norm2}) with \( \alpha =0 \) and \( \beta =\sqrt{\frac{n+2}{n}} \),
one finds that it is in fact smaller than \( \alpha _{c}=1 \) (for \( n=6 \)
and \( a=\frac{1}{2} \), appropriate to a compactification of IIA superstring
theory). But the potential of the truncated theory turns out to be negative,
such that there cannot be accelerating attractor solutions! We would have needed
a positive potential.

So far no example has been found, where a consistent truncation leads to a theory
that admits solutions with an arbitrary amount of accelerating expansion, by
providing an accelerating power-law attractor for the FLRW scale factor.

\section{Discussion\label{sec. discuss}}

Starting from the Einstein-Hilbert action for pure gravity in \( (d+n) \) dimensions,
and also from the bosonic part of various supergravity actions that describe
string/M-theory in the low energy limit, a dimensional reduction down to \( d \)
dimensions has been performed. The \( n \)-dimensional internal space was realised
as a product space, and the moduli fields kept in this reduction determined
the volume of each of the factor spaces. Otherwise these factor spaces have
been considered to be arbitrary but fixed \( d \)-independent Einstein spaces.
The effective \( d \)-dimensional field theory has been studied in Einstein
conformal frame in order to facilitate an easy interpretation of the gravity
results from the lower dimensional point of view. The main focus of this paper
was the investigation of the potential for the scalar fields, the volume moduli,
resulting from the compactification. We were interested in cosmological applications,
not so much in phenomenological aspects but rather in a question of principle:
whether it is possible to find FLRW solutions, from gravity or string/M-theory
compactifications, in which the scale factor exhibits arbitrary amounts of accelerating
expansion, or inflation. 

We have argued that the discussion for the case of multiple scalar fields can
be understood effectively in terms of a single scalar field. The time evolution
essentially picks out, far from the origin in field space, a dominant direction
of the potential. The sum of various exponentials thereby is reduced to a single
effective exponential term. For this term, it is possible to calculate a characteristic
exponent that directly classifies the existing attractor solutions for the scale
factor according to table \ref{tab. relations}. In this way, the intuition
from the single scalar case can be carried over to the more complicated case. 

The calculation of the characteristic exponents for the potentials, arising
from the \( n \)-dimensional geometry and from various fluxes that were realised
as volume forms on certain subspaces, revealed that they are always greater
than the critical exponent in \( d \)-dimensions. The conclusions drawn from
this fact depend on the possible values \( k=0,\pm 1 \) of the curvature of
the spatial sections in the final cosmology. In the flat case with \( k=0 \),
this means that there do not exist any accelerating attractor solutions for
the scale factor. The only way in which accelerating expansion can be obtained
is via the mechanism in the original solutions (subsequent to \cite{ToWo03},
see the introduction) as identified in \cite{EmGa03}. Any such solution can
only have a short phase of accelerating expansion, where the scalar fields are
nearly at rest at some point in field space, around which the potential dominates.
But this phase ends soon as the decelerating attractor is approached. (Note
that we do not claim that there is no way to improve on the \( \mathcal{O}(1) \)
expansion in these scenarios. But the generic existence of decelerating attractors
does make it extremely implausible to achieve sixty e-foldings). For hyperbolic
spatial curvature \( k=-1 \), characteristic exponents greater than \( \alpha _{c} \)
mean that the attractors are exactly critical and non-accelerating, \( a(t)\sim t \).
In this case one can obtain eternal accelerating expansion. This happens in
solutions in which trajectories exist with a point where the scalar fields are
approximately at rest (as above). Then the scale factor starts off accelerating
and, as it approaches the critical attractor, this acceleration does not stop
although it tends to zero. Such a scenario has been identified in \cite{CHNOW03}
by perturbing around the attractor solution. We have not considered the case
\( k=+1 \) because it does not admit attractor solutions. At best, solutions
may behave, for a while, in a similar way as do the \( k=0 \) ones, as suggested
by the analysis of \cite{Hal87}, but they all end in a big crunch. 

In summary, it has been shown that product space compactifications do not lead
to standard inflationary scenarios (with power-law inflation of the scale factor).
Cosmological solutions in this setup do only admit a short phase of accelerating
expansion for \( k=0 \) (and \( k=+1 \)). For \( k=-1 \), the generic solution
can lead to eternal accelerating expansion, but the acceleration quickly tends
to zero. Thus, in order to obtain inflation from higher dimensions and, in particular,
from string/M-theory, it seems to be necessary to think about more complex scenarios
of dimensional reduction. Possibilities of improvement are the inclusion of
further warp factors, possibly the mechanism of consistent truncation presented
in the preceding section, or generically more complicated internal space geometries.
It might also be necessary to consider non-trivial contributions due to Chern-Simons
terms in the supergravity actions. We leave these avenues for future research.

\begin{acknowledgments}

MNRW thanks Paul K. Townsend for very helpful discussions,
and also Nobuyoshi Ohta and John E. Wang. He gratefully acknowledges financial
support from the Gates Cambridge Trust and wishes to thank Trinity
College Cambridge and the organisers of Strings 2003 for enabling
him to visit this conference in Kyoto. 

\end{acknowledgments}

\end{document}